\documentclass[10pt,onecolumn,twoside]{article}
\usepackage{amsmath,graphicx}
\usepackage{amssymb}
\usepackage{amsmath}
\usepackage{cite}
\usepackage{amsthm}
\usepackage{float}
\usepackage{fullpage}

\def\defn{\,\triangleq\,}

\def\gbf{{\mathbf{g}}}
\def\ubf{{\mathbf{u}}}
\def\xbf{{\mathbf{x}}}
\def\ybf{{\mathbf{y}}}
\def\zbf{{\mathbf{z}}}
\def\ebf{{\mathbf{e}}}

\def\xbfhat{{\widehat{\mathbf{x}}}}

\def\Hbf{{\mathbf{H}}}
\def\Dbf{{\mathbf{D}}}
\def\Wbf{{\mathbf{W}}}
\def\Abf{{\mathbf{A}}}
\def\Ibf{{\mathbf{I}}}

\def\Ccal{\mathcal{C}}
\def\Rcal{\mathcal{R}}
\def\Dcal{\mathcal{D}}
\def\Hcal{\mathcal{H}}
\def\Tcal{\mathcal{T}}

\def\R{\mathbb{R}}

\def\argmin{\mathop{\mathrm{arg\,min}}}
\def\prox{\mathrm{prox}}
\def\grad{\nabla}
\def\subgrad{\tilde{\nabla}}

\theoremstyle{definition}
\newtheorem{proposition}{Proposition}

\begin{document}


\title{Parallel proximal methods for total variation minimization}


\author{Ulugbek~S.~Kamilov
\thanks{U.~S.~Kamilov (email: kamilov@merl.com) is with Mitsubishi Electric Research Laboratories, 201 Broadway, Cambridge, MA 02139, USA}}

\maketitle


\begin{abstract}
Total variation (TV) is a widely used regularizer for stabilizing the solution
of ill-posed inverse problems. In this paper, we propose
a novel proximal-gradient algorithm for minimizing TV regularized least-squares 
cost functional. Our method replaces the standard proximal step of TV 
by a simpler alternative that computes several independent proximals. 
We prove that the proposed parallel proximal
method converges to the TV solution, while requiring no sub-iterations.
The results in this paper could enhance the applicability of TV for solving very large scale imaging 
inverse problems.
\end{abstract}


\section{Introduction}
\label{Sec:Intro}

The problem of estimating an unknown signal from noisy linear observations is fundamental
in signal processing. The estimation task is often formulated as the linear inverse problem
\begin{equation}
\label{Eq:LinearInverseProblem}
\ybf = \Hbf \xbf + \ebf,
\end{equation}
where the goal is to compute the unknown signal $\xbf \in \R^N$ from the noisy 
measurements $\ybf \in \R^M$. Here, the matrix $\Hbf \in \R^{M \times N}$ models
the response of the acquisition device and the vector $\ebf \in \R^M$ represents
the measurement noise, which is often assumed to be i.i.d.\ Gaussian. When
the problem~\eqref{Eq:LinearInverseProblem} is ill-posed, the standard approach
is to rely on the regularized least-squares estimator
\begin{equation}
\label{Eq:RegularizedLeastSquareMinimization}
\xbfhat = \argmin_{\xbf \in \R^N} \left\{\frac{1}{2}\|\ybf - \Hbf\xbf\|_{\ell_2}^2 + \Rcal(\xbf)\right\},
\end{equation}
where the functional $\Rcal$ is a regularizer that promotes solutions with desirable properties
such as transform-domain sparsity or positivity.

One of the most widely used regularizers in imaging is the total variation (TV), whose anisotropic variant can be defined as
\begin{equation}
\label{Eq:TotalVariationRegularizer}
\Rcal(\xbf) 
\defn \lambda \|[\Dbf \xbf]\|_{\ell_1}
= \lambda \sum_{n=1}^N \sum_{d = 1}^D |[\Dbf_d \xbf]_n|,
\end{equation}
where $\Dbf: \R^{N} \rightarrow \R^{N \times D}$ is the discrete gradient operator, 
$\lambda > 0$ is a parameter controlling amount of regularization, and $D$ is
the number of dimensions in the signal. The matrix $\Dbf_d$ 
denotes the finite difference operation along the dimension $d$ with appropriate boundary
conditions (periodization, etc.). The TV prior has 
been originally introduced by Rudin \textit{et al.}~\cite{Rudin.etal1992} as a 
regularization approach capable of removing noise, while preserving image edges. 
It is often interpreted as a sparsity-promoting $\ell_1$-penalty on the image gradient. TV regularization has proven to be 
successful in a wide range of applications in the context of sparse recovery of images 
from incomplete or corrupted measurements~\cite{Bronstein.etal2002, 
Afonso.etal2010, Candes.etal2006, Lustig.etal2007, Louchet.Moisan2008, Oliveira.etal2009, Kamilov.etal2015}.

The minimization~\eqref{Eq:RegularizedLeastSquareMinimization} with the TV regularization 
is a nontrivial optimization task. The challenging aspects are the non-smooth nature
of the regularization term~\eqref{Eq:TotalVariationRegularizer} and the massive quantity of data 
that typically needs to be processed. Proximal gradient 
methods~\cite{Bauschke.Combettes2010} such as
iterative shrinkage/thresholding algorithm (ISTA)~\cite{Figueiredo.Nowak2003, Bect.etal2004, Daubechies.etal2004, Bioucas-Dias.Figueiredo2007, Beck.Teboulle2009} or 
alternating direction method of multipliers (ADMM)~\cite{Eckstein.Bertsekas1992, Boyd.etal2011, Qin.etal2011} 
are standard 
approaches to circumvent the non-smoothness of the TV regularizer. 

For the optimization
problem~\eqref{Eq:RegularizedLeastSquareMinimization}, ISTA can be written as
\begin{subequations}
\label{Eq:ISTA}
\begin{align}
\label{Eq:ISTAa}
&\zbf^t \leftarrow \xbf^{t-1} - \gamma_t \Hbf^T \left(\Hbf\xbf^{t-1}-\ybf\right)\\
\label{Eq:ISTAb}
&\xbf^t \leftarrow \prox_{\gamma_t \Rcal} (\zbf^t),
\end{align}
\end{subequations}
where $\gamma_t > 0$ is a step-size that can be determined a priori to ensure 
convergence~\cite{Beck.Teboulle2009}. Iteration~\eqref{Eq:ISTA} combines 
the gradient-descent step~\eqref{Eq:ISTAa} with a proximal operation~\eqref{Eq:ISTAb}
defined as
\begin{equation}
\label{Eq:ProximalOperator}
\prox_{\gamma \Rcal}(\zbf) \defn \argmin_{\xbf \in \R^N} \left\{\frac{1}{2}\|\xbf-\zbf\|_{\ell_2}^2 + \gamma \Rcal(\xbf)\right\}.
\end{equation}
The proximal operator corresponds to the regularized solution of the denoising problem 
where $\Hbf$ is an identity. Because of its simplicity, ISTA and its accelerated variants 
are among the methods of choice for solving practical linear inverse 
problems~\cite{ Bioucas-Dias.Figueiredo2007, Beck.Teboulle2009}.
Nonetheless, ISTA--based optimization of TV is complicated by the
fact that the corresponding proximal operator does not admit a closed form solution.
Practical implementations 
rely on computational solutions that require an additional nested optimization algorithm for 
evaluating the TV proximal~\cite{Beck.Teboulle2009a, Qin.etal2011}. 
This typically leads to a prohibitively slow reconstruction when dealing with
very large scale imaging problems such as the ones in 3D microscopy~\cite{Kamilov.etal2015}.

In this paper, we propose a novel approach for solving TV--based imaging problems
that requires no nested iterations. This is achieved by substituting the proximal of
TV with an alternative that amounts to evaluating several simpler proximal operators.
One of our major contributions is the proof that the approach can achieve the true TV 
solution with arbitrarily high precision.
We believe that the results presented in this paper are useful to practitioners working with
very large scale problems that are common in 3D imaging, where the 
bottleneck is often in the evaluation of the TV proximal.


\section{Main Results}
\label{Sec:MainResult}

In this section, we present our main results.
We start by introducing the proposed approach and then follow up by analyzing 
its convergence.

\subsection{General formulation}

We turn our attention to a more general optimization problem
\begin{equation}
\xbfhat = \argmin_{\xbf \in \R^N} \left\{\Ccal(\xbf)\right\},
\end{equation}
where the cost functional is of the following form
\begin{equation}
\label{Eq:CostFunctional}
\Ccal(\xbf) = \Dcal(\xbf) + \Rcal(\xbf) = \Dcal(\xbf) + \frac{1}{K} \sum_{k = 1}^K \Rcal_k(\xbf).
\end{equation}
The precise connection between~\eqref{Eq:CostFunctional} and TV-regularized 
cost functional will be discussed shortly.
We assume that the data-fidelity term $\Dcal$ is convex and differentiable
with a Lipschitz continuous gradient. This means that there exists a constant $L > 0$
such that, for all $\xbf, \zbf \in \R^N$, $\|\grad \Dcal(\xbf) - \Dcal(\zbf) \|_{\ell_2} \leq L \|\xbf -\zbf\|_{\ell_2}$. We also assume that each $\Rcal_k$ is a continuous, convex function that is 
possibly nondifferentiable and that the optimal value $\Ccal^\ast$ is finite
and attained at $\xbf^\ast$.

We consider parallel proximal algorithms that have the following form
\begin{subequations}
\label{Eq:ParallelProximalAlgorithm}
\begin{align}
\label{Eq:ParallelProximalAlgorithm1}
&\zbf^t \leftarrow \xbf^{t-1} - \gamma_t \grad \Dcal(\xbf^{t-1}) \\
\label{Eq:ParallelProximalAlgorithm2}
&\xbf^t \leftarrow \frac{1}{K} \sum_{k = 1}^K \prox_{\gamma_t \Rcal_k}(\zbf^t),
\end{align}
\end{subequations}
where $\prox_{\gamma_t \Rcal_k}$ is the proximal operator associated with $\gamma_t \Rcal_k$.
We are specifically interested in the case where the proximals $\prox_{\gamma_t \Rcal_k}$ 
have a closed form, in which case they are preferable to the computation of the full
proximal $\prox_{\gamma_t \Rcal}$.

We now establish a connection between~\eqref{Eq:CostFunctional} and TV-regularized cost. 
Define a linear transform 
$\Wbf: \R^N \rightarrow \R^{N \times D \times 2}$ that consists of two sub-operators:
the averaging operator $\Abf: \R^N \rightarrow \R^{N \times D}$ and the discrete gradient 
$\Dbf$ as in~\eqref{Eq:TotalVariationRegularizer}, both normlized by $1/\sqrt{2}$. 
The averaging operator consists of $D$ matrices $\Abf_d$ 
that denote the pairwise averaging along the dimension $d$. Accordingly, the operator $\Wbf$ is a
union of scaled and shifted discrete Haar wavelet and scaling functions along each
dimension~\cite{Mallat2009}. Since we consider all possible shifts along each dimension
the transform is redundant and can be interpreted as the union of $K = 2D$, scaled, orthogonal 
tranforms 
\begin{equation}
\Wbf =
\begin{bmatrix}
\Wbf_1 \\ 
\vdots \\ 
\Wbf_K
\end{bmatrix}.
\end{equation}
The transform $\Wbf$ and its pseudo-inverse 
\begin{equation}
\Wbf^\dagger = \frac{1}{K} [\Wbf_1^T \dots \Wbf_K^T]
\end{equation}
satisfy the following two properties of Parseval frames~\cite{Unser.Tafti2014}
\begin{equation*}
\argmin_{\xbf \in \R^N} \left\{\frac{1}{2}\|\zbf - \Wbf\xbf\|_{\ell_2}^2\right\}  = \Wbf^\dagger \zbf \quad(\text{for all } \zbf \in \R^{KN})
\end{equation*}
and
\begin{equation}
\Wbf^\dagger\Wbf = \Ibf. 
\end{equation}
One can thus express the TV regularizer as the following sum
\begin{equation}
\Rcal(\xbf) = \lambda \sqrt{2} \sum_{k = 1}^K \sum_{n \in \Hcal_k} |[\Wbf_k \xbf]_n|,
\end{equation}
where $\Hcal_k \subseteq [1 \dots N]$ is the set of all detail coefficients of the transform $\Wbf_k$.
Then, the proposed parallel proximal algorithm for TV can be expressed as follows
\begin{subequations}
\label{Eq:ParTV}
\begin{align}
\label{Eq:ParTVa}
&\zbf^t \leftarrow \xbf^{t-1} - \gamma_t \Hbf^T \left(\Hbf\xbf^{t-1}-\ybf\right)\\
\label{Eq:ParTVb}
&\xbf^t \leftarrow \frac{1}{K} \sum_{k = 1}^K \Wbf^T_k \Tcal (\Wbf_k\zbf^t; \sqrt{2}K \gamma_t \lambda ),
\end{align}
\end{subequations}
where $\Tcal$ is the component-wise shrinkage function
\begin{equation}
\label{Eq:IsotropicShrinkage}
\Tcal(y; \tau) \defn \max(|y|-\tau, 0) \frac{y}{|y|},
\end{equation}
which is applied only on scaled differences $\Dbf\zbf^t$.

The algorithm in~\eqref{Eq:ParTV} is closely related to a technique called 
cycle spinning~\cite{Coifman.Donoho1995} that is commonly used 
for improving the performance of wavelet-domain denoising. In particular,
when $\Hbf = \Ibf$ and $\gamma_t = 1$, for all $t = 1, 2, \dots$, the algorithm
yields the solution
\begin{equation}
\xbfhat \leftarrow \Wbf^\dagger \Tcal (\Wbf\ybf; \sqrt{2}K\lambda),
\end{equation}
which can be interpreted as the traditional cycle spinning
algorithm restricted to the Haar wavelet-transform. In the context of image denoising, the connections between TV and 
cycle-spinning were originally established in~\cite{Kamilov.etal2012}.

\subsection{Theoretical convergence}
\label{Sec:TheoreticalConvergence}

The convergence results in this section assume that the gradient of $\Dcal$ and 
subgradients of $\Rcal_k$ are bounded, i.e., there exists $G > 0$ such that for all $k$ and $t$, 
$\|\grad \Dcal(\xbf^t)\|_{\ell_2} \leq G$ and $\|\partial \Rcal_k(\xbf^t)\|_{\ell_2} \leq G$.
The following proposition is crucial for establishing the convergence of the parallel 
proximal algorithm.

\begin{proposition}
\label{Prop:Prop1}
Consider the cost function~\eqref{Eq:CostFunctional} and the algorithm~\eqref{Eq:ParallelProximalAlgorithm}. Then, for all $t = 1, 2, \dots$,
and for any $\xbf \in \R^N$, we have
\begin{align}
\Ccal(\xbf^t) &-\Ccal(\xbf) \\
&\leq \frac{1}{2\gamma_t}\left(\|\xbf^{t-1}-\xbf\|_{\ell_2}^2 - \|\xbf^{t}-\xbf\|_{\ell_2}^2\right) + 8\gamma_t G^2 \nonumber.
\end{align}
\end{proposition}

\noindent
\emph{Proof: } See Appendix.

\vspace{0.25cm}\noindent
Proposition~\ref{Prop:Prop1} allows us to develop various types of convergence results.
For example, if $\xbf^\ast$ is the optimal point and if we pick
a sufficiently small step
$\gamma_t \leq (\Ccal(\xbf^t)-\Ccal(\xbf^\ast))/(8G^2)$, then the $\xbf^t$ will be closer
to $\xbf^\ast$ than $\xbf^{t-1}$. This argument can be formalized into the following
proposition.
\begin{proposition}
\label{Prop:Prop2}
Assume a fixed step size $\gamma_t = \gamma > 0$. Then, we have that
\begin{equation}
\label{Eq:ConvergenceResult}
\liminf_{t \rightarrow \infty} \left(\Ccal(\xbf^t)-\Ccal^\ast\right) \leq 8\gamma G^2.
\end{equation}
\end{proposition}
\noindent
\emph{Proof: } See Appendix.

\vspace{0.25cm}\noindent
Proposition~\ref{Prop:Prop2} states that for a constant step-size, convergence
can be established to the neighborhood of the optimimum, which can be made 
arbitrarily close to $0$ by letting $\gamma \rightarrow 0$.


\section{Experiments}
\label{Sec:Experiments}

\begin{figure}[t]
\centering\includegraphics[width=8.5cm]{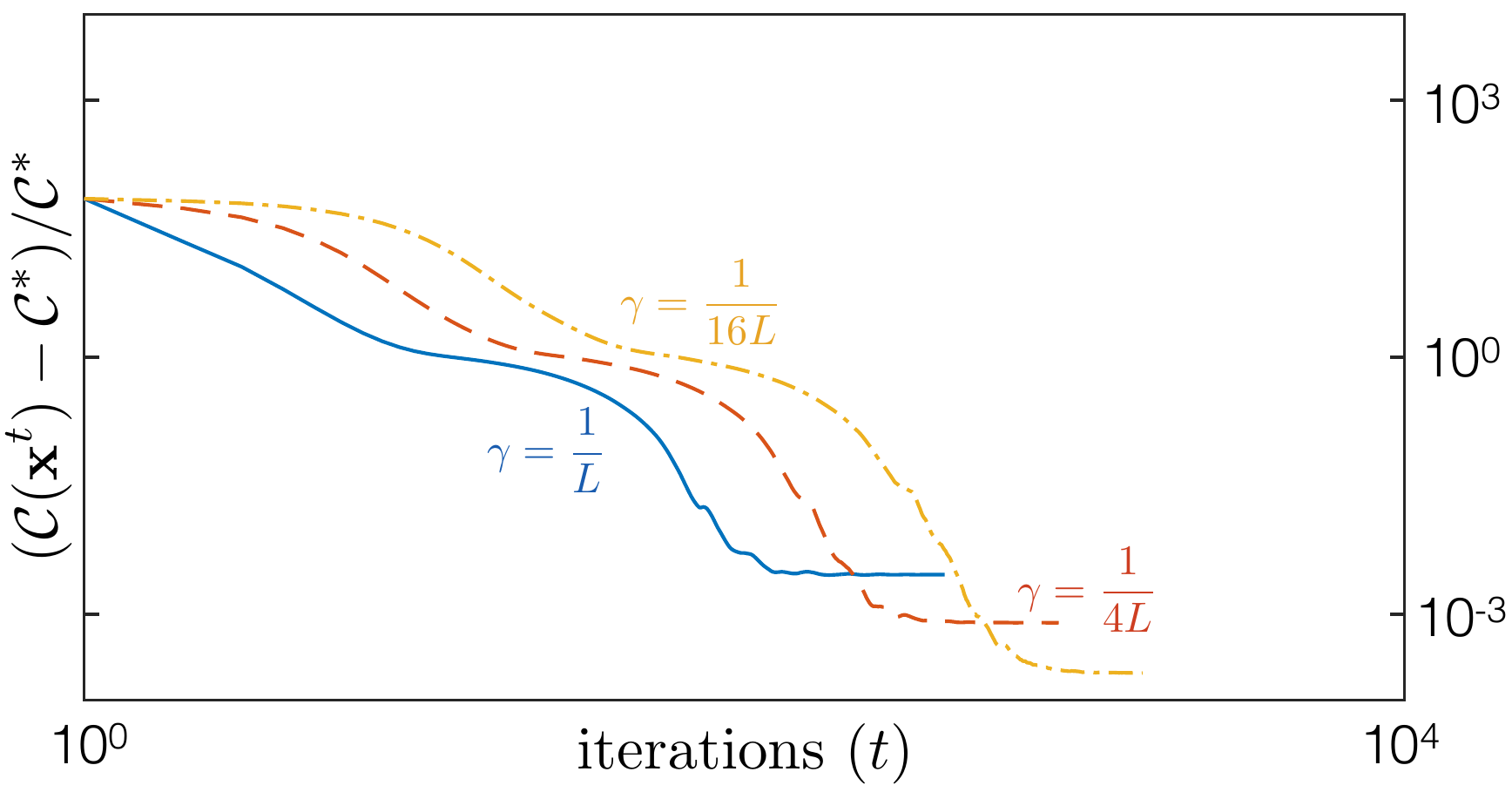}
\caption{Reconstruction of a \textit{Shepp-Logan} phantom from
noisy linear measurements. The relative gap ${(\Ccal(\xbf^t)-\Ccal^\ast)/\Ccal^\ast}$ is plotted against
the iteration number for 3 distinct step-sizes $\gamma$.
The plot illustrates the convergence of the fast parallel proximal algorithm to the minimizer
of the TV cost functional.}
\label{Fig:Fig1}
\end{figure}

\begin{figure}[t]
\centering\includegraphics[width=8.5cm]{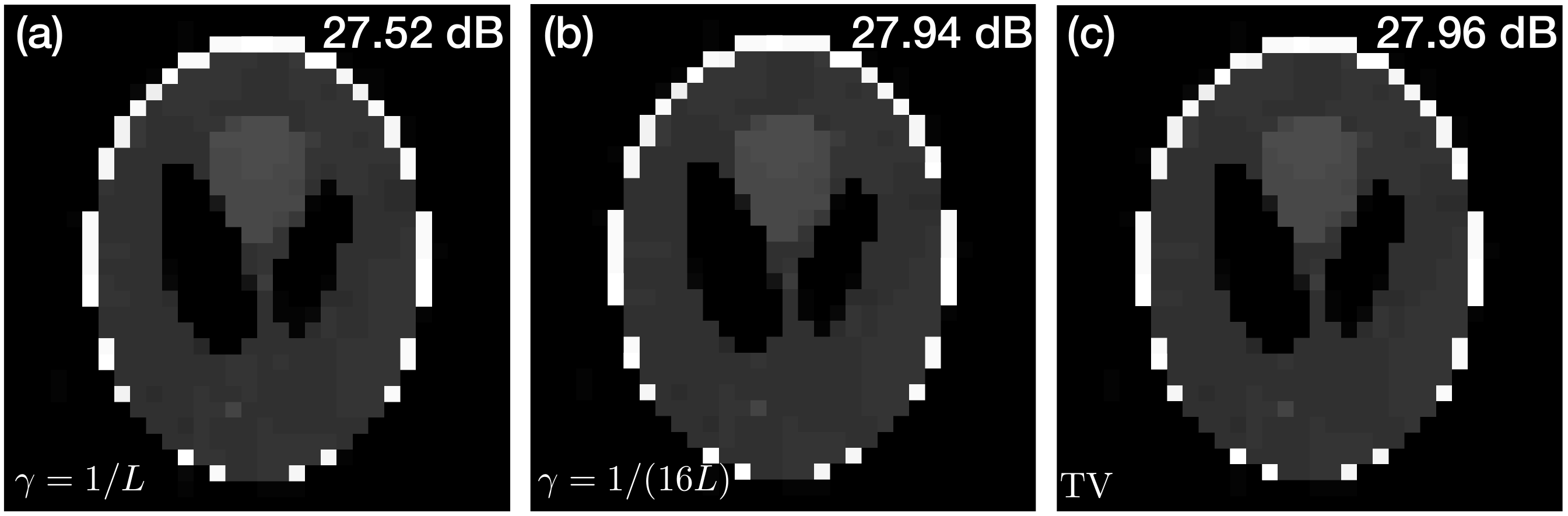}
\caption{Reconstructed \textit{Shepp-Logan} images for (a) $\gamma = 1/L$;
(b) $\gamma = 1/(16L)$; (c) the true TV solution. Even for $\gamma = 1/L$ the solution
of the accelerated parallel proximal algorithm is visually and quantitatively close to the 
true TV result.}
\label{Fig:Fig2}
\end{figure}

In this section, we empirically illustrate that our results hold more generally than suggested by Proposition~\ref{Prop:Prop2}. Specifically, we consider the accelerated parallel proximal
algorithm based on FISTA~\cite{Beck.Teboulle2009} 
\begin{subequations}
\label{Eq:FastParallelProximalAlgorithm}
\begin{align}
\label{Eq:FastParallelProximalAlgorithm1}
&\zbf^t \leftarrow \ubf^{t-1} - \gamma_t \grad \Dcal(\ubf^{t-1}) \\
\label{Eq:FastParallelProximalAlgorithm2}
&\xbf^t \leftarrow \frac{1}{K} \sum_{k = 1}^K \prox_{\gamma_t \Rcal_k}(\zbf^t) \\
\label{Eq:FastParallelProximalAlgorithm3}
&q_{t} \leftarrow (1 + \sqrt{1+4q_{t-1}^2})/2 \\
\label{Eq:FastParallelProximalAlgorithm4}
&\ubf^t \leftarrow \xbf^t + (q_{t-1}-1)/q_t)(\xbf^t-\xbf^{t-1})
\end{align}
\end{subequations}
with $\ubf^0 = \xbf^0$, $q_0 = 1$, and $\gamma_t = \gamma$.
Method~\eqref{Eq:FastParallelProximalAlgorithm} preserves the simplicity of 
the ISTA approach~\eqref{Eq:ParallelProximalAlgorithm} but provides a significantly 
better rate of convergence, which enhances potential applicability of the method. 
We consider an estimation problem 
where the Shepp-Logan phantom of size $32 \times 32$ is reconstructed from $M = 512$ linear
measurements with AWGN corresponding to $30$ dB SNR.
The measurement matrix is i.i.d.\ Gaussian 
$[\Hbf]_{mn} \sim \mathcal{N}(0, 1/M)$.
Figure~\ref{Fig:Fig1} illustrates the per-iteration gap
$(\Ccal(\xbf^t)-\Ccal^\ast)/\Ccal^\ast$, where $\xbf^t$ is computed with the fast parallel proximal 
method~\eqref{Eq:FastParallelProximalAlgorithm}
and $\Ccal$ is the TV-regularized least-squares cost. 
The regularization parameter $\lambda$ was manually selected for the optimal SNR performance
of TV. We compare 3 different step-sizes
$\gamma = 1/L$, $\gamma = 1/(4L)$, and $\gamma = 1/(16L)$, where 
$L = \lambda_{\text{\tiny max}}(\Hbf^T \Hbf)$ is the Lipschitz constant.
Proposition~\ref{Prop:Prop2} suggests that the gap $(\Ccal(\xbf^t)-\Ccal^\ast)$ is proportional
to the step-size and shrinks to 0 as the step-size is reduced. Such behavior is clearly observed
in Figure~\ref{Fig:Fig1}, which suggests that our results potentially
hold for the accelerated parallel proximal algorithm. Figure~\ref{Fig:Fig2}
compares the quality of the estimated images, for $\gamma = 1/L$ and $\gamma = 1/(16L)$,
against the TV solution. We note that, even for $\gamma = 1/L$, the solution of our algorithm 
is very close to the true TV result, both qualitatively and quantitatively. This implies that,
while requiring no nested iterations, 
our parallel proximal approach can potentially approximate the solution of TV with 
arbitrarily accurate precision at $\mathcal{O}(1/t^2)$ convergence rate of FISTA.


\section{Relation to Prior Work}

The results in this paper are most closely related to the work on TV--based imaging 
by Beck and Teboulle~\cite{Beck.Teboulle2009a}. While their approach requires
additional nested optimization to compute the TV proximal, we avoid this by relying on
multiple simplified proximals computed in parallel. Our proofs rely on several results
from convex optimization that were used by Bertsekas~\cite{Bertsekas2011} for analyzing a 
different family of algorithms called incremental proximal methods. Finally, two
earlier papers with the author describe the relationship between cycle spinning and TV~\cite{Kamilov.etal2012, Kamilov.etal2014a}, but concentrate on a fundamentally different 
families of optimization algorithms.


\section{Conclusion}
\label{Sec:Conclusion}

The parallel proximal method and its accelerated version, which were presented 
in this paper, are beneficial in the context of TV regularized image reconstruction, especially
when  the computation of the TV proximal is costly. We presented a combination of theoretical and 
empirical evidence demonstrating that these methods can compute the TV solution 
at the competitive global convergence rates without
resorting to expensive sub-iterations. Future work will aim at extending the theoretical
analysis presented here and by applying the methods to a larger class
of imaging problems.


\section{Appendix}
\label{Sec:Appendix}

We now prove the propositions in Section~\ref{Sec:TheoreticalConvergence}. 
The formalism used here is closely related to the analysis of 
incremental proximal methods that were studied by Bertsekas~\cite{Bertsekas2011}.
Related techniques were also used to analyze the convergence of recursive cycle spinning
algorithm in~\cite{Kamilov.etal2014a}.

\subsection{Proof of Proposition~\ref{Prop:Prop1}}

We define an intermediate quantity $\xbf_k^t \defn \prox_{\gamma_t \Rcal_k}(\zbf^t)$.
The optimality conditions for~\eqref{Eq:ParallelProximalAlgorithm2} imply that there must
exist $K$ subgradient vectors $\subgrad \Rcal_k(\xbf_k^t) \in \partial \Rcal_k(\xbf_k^t)$
such that
\begin{equation}
\label{Eq:GradFormulation1}
\xbf_k^t = \xbf^{t-1}-\gamma_t \left(\grad \Dcal(\xbf^{t-1}) + \subgrad \Rcal_k(\xbf_k^t)\right).
\end{equation}
This implies that 
\begin{equation}
\label{Eq:GradFormulation2}
\xbf^t = \xbf^{t-1} - \gamma_t \left(\grad \Dcal(\xbf^{t-1}) + \gbf^t\right),
\end{equation}
where 
\begin{equation*}
\gbf^t \defn \frac{1}{K} \sum_{k = 1}^K \subgrad \Rcal_k(\xbf_k^t).
\end{equation*}
Then we can write
\begin{align}
\label{Eq:MainInequality}
\|&\xbf^t- \xbf\|_{\ell_2}^2 = \|\xbf^{t-1} - \gamma_t (\grad \Dcal(\xbf^{t-1}) + \gbf^t) -\xbf\|_{\ell_2}^2 \\
& =  \|\xbf^{t-1} - \xbf\|_{\ell_2}^2 - 2\gamma_t\langle \grad \Dcal(\xbf^{t-1}) + \gbf^t, \xbf^{t-1}-\xbf\rangle \nonumber\\
&\quad\quad+\gamma_t^2 \|\grad \Dcal(\xbf^{t-1})+\gbf^t\|_{\ell_2}^2 \nonumber
\end{align}
By using the triangle inequality and noting that all the subgradients are bounded, we can bound the last term as
\begin{equation}
\label{Eq:Bound1}
\|\grad \Dcal(\xbf^{t-1})+\gbf^t\|_{\ell_2}^2 \leq 4G^2.
\end{equation}
To bound the second term we proceed in two steps. We first write that
\begin{align}
\label{Eq:Bound2}
\langle \grad \Dcal(\xbf^{t-1}), &\,\xbf^{t-1}- \xbf \rangle \geq \Dcal(\xbf^{t-1}) - \Dcal(\xbf) \\
&\geq \Dcal(\xbf^t) - \langle \grad \Dcal(\xbf^t), \xbf^t - \xbf^{t-1}\rangle- \Dcal(\xbf) \nonumber\\
&\geq \Dcal(\xbf^t) - \Dcal(\xbf) - 2\gamma_t G^2 \nonumber,
\end{align}
where we used the convexity of $\Dcal$, the Cauchy-Schwarz inequality, and the bound
on the gradients. In a similar way, we can write that
\begin{align}
\label{Eq:Bound3}
\langle \gbf^t, &\, \xbf^{t-1}-\xbf \rangle = \frac{1}{K} \sum_{k = 1}^K \langle \subgrad \Rcal_k(\xbf_k^t), \xbf^{t-1}-\xbf \rangle \\
& \geq \frac{1}{K} \sum_{k = 1}^K (\Rcal_k(\xbf_k^t) - \Rcal_k(\xbf)) - 2\gamma_t G^2 \nonumber\\
& \geq \Rcal(\xbf^t) - \Rcal(\xbf) - 4\gamma_t G^2, \nonumber
\end{align}
where we used the convexity of $\Rcal_k$s, the relationships~\eqref{Eq:GradFormulation1} 
and~\eqref{Eq:GradFormulation2}, as well as bounds obtained via the 
Cauchy-Schwarz inequality.
By plugging~\eqref{Eq:Bound1},~\eqref{Eq:Bound2}, and~\eqref{Eq:Bound3} 
into~\eqref{Eq:MainInequality} and by reorganizing the terms, we obtain the claim.

\subsection{Proof of Proposition~\ref{Prop:Prop2}}

By following an approach similar to Bertsekas~\cite{Bertsekas2011}, 
we prove the result by contradiction. 
Assume that~\eqref{Eq:ConvergenceResult} does not hold. Then, there must
exist $\epsilon > 0$ such that
\begin{equation}
\liminf_{t \rightarrow \infty} (\Ccal(\xbf^t) - \Ccal^\ast) > 8\gamma G^2 + 2\epsilon
\end{equation}
Let $\bar{\xbf} \in \R^N$ be such that
\begin{equation}
\label{Eq:LowerBoundXBar}
\liminf_{t \rightarrow \infty} \Ccal(\xbf^t) - 8\gamma G^2 - 2\epsilon \geq \Ccal(\bar{\xbf})
\end{equation}
and let $t_0$ be large enough so that for all $t \geq t_0$, we have
\begin{equation}
\label{Eq:Limit}
\left|\Ccal(\xbf^t) - \liminf_{t \rightarrow \infty} \Ccal(\xbf^t)\right| \leq \epsilon.
\end{equation}
By combining~\eqref{Eq:LowerBoundXBar} and~\eqref{Eq:Limit}, we obtain
that for all $t \geq t_0$
\begin{equation}
\Ccal(\xbf^t) - \Ccal(\bar{\xbf}) \geq 8\gamma G^2 + \epsilon.
\end{equation}
Then from Proposition~\ref{Prop:Prop1}, for all $t \geq t_0$,
\begin{align}
\|\xbf^t - &\bar{\xbf}\|_{\ell_2}^2 \nonumber\\
&\leq \|\xbf^{t-1}-\bar{\xbf}\|_{\ell_2}^2 - 2\gamma (\Ccal(\xbf^t)-\Ccal(\bar{\xbf})) + 16 \gamma^2 G^2 \nonumber\\
&\leq \|\xbf^{t-1}-\bar{\xbf}\|_{\ell_2}^2- 2\gamma \epsilon.
\end{align}
By iterating the inequality over $t$, we have for all $t \geq t_0$,
\begin{equation}
\|\xbf^t - \bar{\xbf}\|_{\ell_2}^2 \leq \|\xbf^{t_0} - \bar{\xbf}\|_{\ell_2} - 2(t-t_0) \gamma \epsilon,
\end{equation}
which cannot hold for arbitrarily large $t$. This completes the proof.


\bibliographystyle{IEEEtran}


\end{document}